\documentclass[conference]{IEEEtran}
\usepackage{cite}
\usepackage{amsmath,amssymb,amsfonts}
\usepackage{graphicx}
\usepackage{textcomp}
\usepackage{xcolor}
\usepackage{cleveref}
\usepackage[utf8]{inputenc}
\usepackage{url}
\usepackage{subcaption}
\usepackage{showexpl}
\usepackage{algpseudocode}
\usepackage{algorithm}
\usepackage{algorithmicx}
\usepackage{algcompatible}
\usepackage[english]{babel}

\definecolor{antiquebrass}{rgb}{0.8, 0.58, 0.46}
\definecolor{airforceblue}{rgb}{0.36, 0.54, 0.66}
\definecolor{tyrianpurple}{rgb}{0.4, 0.01, 0.24}


\title{Enabling Autonomous Electron Microscopy for Networked Computation and Steering
\thanks{This research is sponsored in part by the INTERSECT Initiative as part of the Laboratory Directed Research and Development Program and in part by RAMSES project of Advanced Scientific Computing Research program, U.S. Department of Energy, and in part by the Office of Basic Energy Sciences, Division of Materials Sciences and Engineering, U.S. Department of Energy, and is performed at Oak Ridge National Laboratory managed by UT-Battelle, LLC for U.S. Department of Energy under Contract No. DE-AC05-00OR22725.
The United States Government retains and the publisher, by accepting the article for publication, acknowledges that the United States Government retains a nonexclusive, paid-up, irrevocable, world-wide license to publish or reproduce the published form of this manuscript, or allow others to do so, for United States Government purposes. The Department of Energy will provide public access to these results of federally sponsored research in accordance with the DOE Public Access Plan (http://energy.gov/downloads/doe-public-access-plan).
}
}

\author{
\IEEEauthorblockN{
Anees Al-Najjar, Nageswara S. V. Rao, Ramanan Sankaran\\
Maxim Ziatdinov, Debangshu Mukherjee, Olga Ovchinnikova}
\IEEEauthorblockA{\textit{Computational Sciences and Engineering Division}\\
\textit{Oak Ridge National Laboratory}\\ Oak Ridge, TN, USA\\
\{alnajjaram,raons,sankaranr,ziatdinovma,mukherjeed,ovchinnikovo\}@ornl.gov}
\and
\IEEEauthorblockN{
Kevin Roccapriore,\\ Andrew R. Lupini, Sergei V. Kalinin}
\IEEEauthorblockA{
\textit{Center for Nanophase Materials Science}\\
\textit{Oak Ridge National Laboratory}\\ Oak Ridge, TN, USA\\
\{roccapriorkm,arl1000,kalininsv\}@ornl.gov}
}

\begin{document}

\maketitle

\begin{abstract}
Advanced electron microscopy workflows require an ecosystem of microscope instruments and computing systems possibly located at different sites to conduct remotely steered and automated experiments.
Current workflow executions involve manual operations for steering and measurement tasks, which are typically performed from control workstations co-located with microscopes;  consequently, their operational tempo and effectiveness are limited.
We propose an approach based on separate data and control channels for such an ecosystem of Scanning Transmission Electron Microscopes (STEM) and computing systems, for which no general solutions presently exist, unlike the neutron and light source instruments. We demonstrate automated measurement transfers and remote steering of Nion STEM physical instruments over site networks. We propose a Virtual Infrastructure Twin (VIT) of this ecosystem, which is used to develop and test our steering software modules without requiring access to the physical instrument infrastructure. Additionally, we develop a VIT for a multiple laboratory scenario, which illustrates the applicability of this approach to ecosystems connected over wide-area networks, for the development and testing of software modules and their later field deployment. 
\end{abstract}



\section{introduction}
\label{sec:introduction}

There is an increasing interest in scientific workflows that incorporate remotely controlled, automated experiments over collections of physical instruments and computing systems. 
Often, these resources are located at geographically dispersed sites, and they need to be federated over wide-area networks to form \textit{ecosystems} that seamlessly support these workflows \cite{Lu2020,9307775}.
Recent workflow developments enable the use of Artificial Intelligence (AI) codes both as a part of scientific computations and data analyses, and for orchestrating automated experiments at potentially remote physical instruments.
In particular, the latter tasks may involve configuring instruments, collecting and transferring measurements and analyzing them to extract parameters for the next set of remote experiments.
Effective execution of such workflows requires the application codes to be customized for remote computing and storage resources connected over networks. 
Currently, science users manually orchestrate several of these tasks, which may have to be repeated with different parameters based on analyses results. 
These human-driven, time-consuming processes limit the scalability and execution tempo of scientific workflows, and often lead to inefficient idling of expensive resources. 
 
\begin{figure}
\centering
\includegraphics[width=0.5\textwidth]{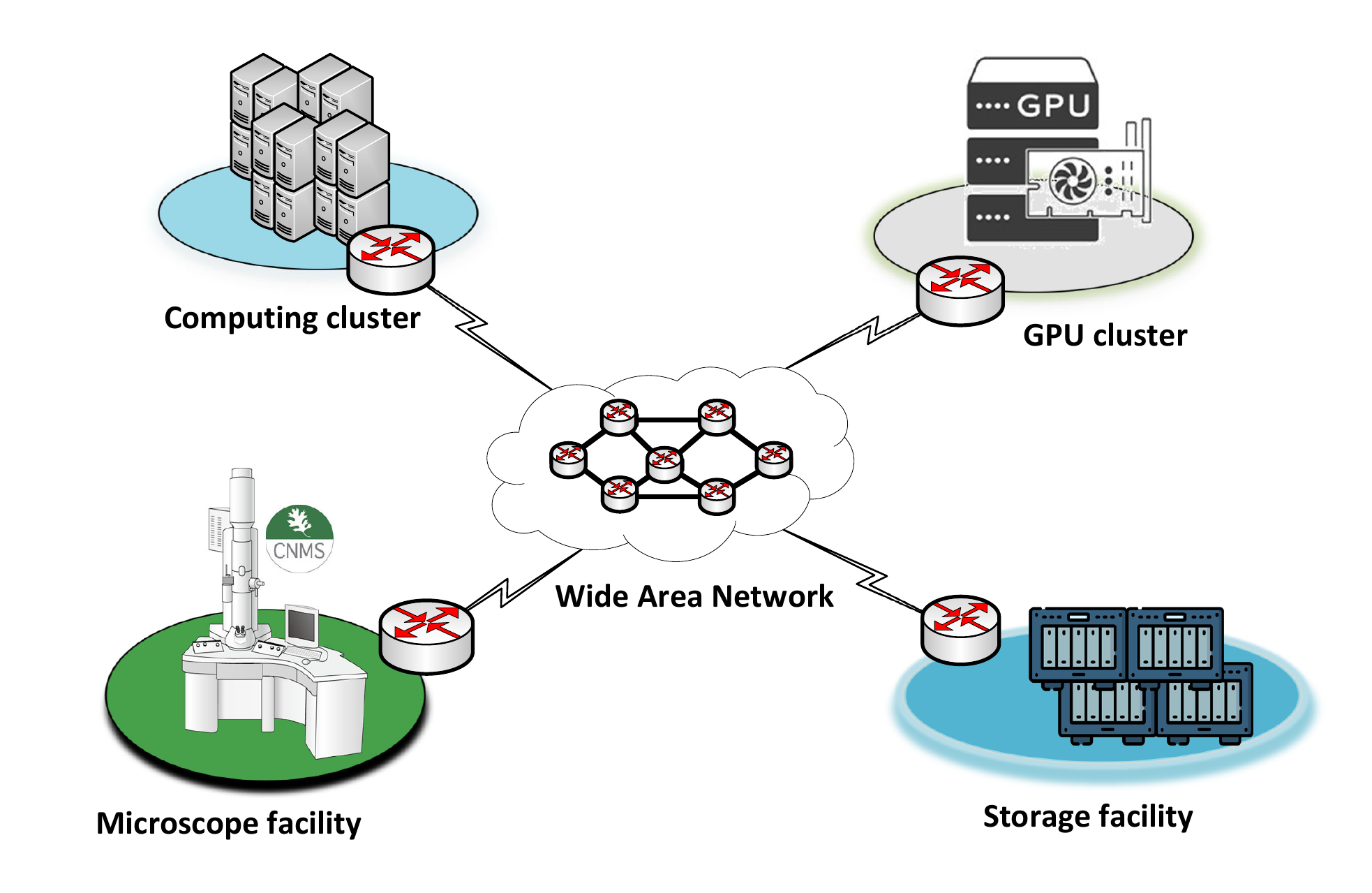}
\caption{\small An ecosystem of microscope, computing and storage sites connected over a wide-area network is needed to seamlessly support science workflows that require automated and remotely controlled experiments.} 
\label{fig:microscopy_federation}
\vspace*{-0.2in}
\end{figure}

The electron microscopy workflows are expected to significantly benefit from the computing and storage capabilities provided by these ecosystems \cite{nmat_dataEM_perspective,levental2021ultrafast}; general versions of such science ecosystems may utilize diverse instruments, for example, light sources \cite{bicer2017real}. 
We consider an ecosystem of Scanning Transmission Electron Microscopes (STEM) that are extensively used in science workflows \cite{4dstem_review}, for example, ptychography using atomic imaging for novel materials synthesis.
More generally, the transmission electron microscope with electron imaging, electron diffraction, and spectroscopy capabilities is aptly called  ``A Synchrotron in a Microscope''\cite{synchrotron_tem} and has extensive uses in physical and life sciences \cite{kirkland_book}.
Currently, no general software frameworks exist to build these microscopy ecosystems, unlike others  such as the Experimental Physics and Industrial Control System (EPICS) \cite{epics} widely deployed at neutron and light source facilities.
The microscopes are typically controlled by local control computers using custom Windows software, and the computing systems are typically Linux platforms located in different networks separated by firewalls.
To fully realize the potential of STEM ecosystems,  \textit{eSolutions} for software design, testing and implementation, are needed for seamlessly collecting and transferring measurements and steering the microscopes from remote computing systems, as illustrated in Figure~\ref{fig:microscopy_federation}.

We consider Nion STEM systems at Oak Ridge National Laboratory (ORNL) and remote Linux computing systems with GPUs to support measurements and steering, in addition to computations that use the measurements. 
Our contributions are two-fold: (a) demonstration of remote steering and automated measurement transfers over the ORNL STEM ecosystem (more than two decades after some of these microscopes were installed), and (b) Virtual Infrastructure Twins (VIT) that support the development and testing of ecosystem software prior to and in preparation for deployment. 
In prior work, the measurement transfers required manual steps, and the steering was only available from the control computer co-located with the instrument.
For part (a), the measurements are made available at remote Linux workstations using a Network Attached Storage (NAS) at the instrument site, and remote mounting its file system on the servers. We  develop Pyro server and client codes to enable remote microscope commands to be sent and executed on the microscope using the Nion Swift microscope software. 
For part (b), we developed VITs of ORNL and multi-site ecosystems with network control channels between remote computing systems and microscope systems, and tested initial Pyro codes prior to deployment(in part (a)).

The development and testing of the ecosystem software modules (such as Pyro server client codes) typically takes several days or longer. It is not cost-effective to require physical access to the microscope for the entire duration, and indeed may not be necessary. 
Instead, we develop a VIT of the ORNL ecosystem that emulates the network and computing systems, and incorporates the Nion Swift simulator.
It provides the software environment nearly identical to the deployed system,
and is used to develop and test Pyro server and client codes, without tying up an expensive microscope and its human operator for several days.
We also develop another VIT of four laboratory sites connected over a wide-area network, and demonstrate the remote steering capability across the sites.
It illustrates the broader applicability of the VIT approach for (initial) development of ecosystem software components without requiring access to the physical infrastructure.

The organization of this paper is as follows.
A brief account of STEM and the associated scientific workflows and ORNL infrastructure are presented
in Section \ref{sec:Electron Microscopy}.
Design and implementation of data and control channels are presented in Section \ref{sec:design_and_implementation}.
Experimental results over ORNL STEM ecosystem are presented in \ref{sec:experimental_setup}.
VITs of ORNL  site and 
four-sites scenario are described in Section \ref{sec:Virtual Infrastructure Twin}.
Conclusions and directions for future research are presented in
Section~\ref{sec: conclusion}.


\section{STEM Workflows and Infrastructure}
\label{sec:Electron Microscopy}

Scientific STEM workflows are supported by an infrastructure of instruments and their control computers typically connected over local networks.

\subsection{STEM Systems, Workflows and Software}


Scientific workflows that utilize STEM are quite varied and extensive.

\subsubsection{Principles and applications}
The STEM images are generated by scanning a focused electron beam across a thin sample. The distribution of transmitted (and/or scattered) electrons in the detector plane depends on the sample composition and structure. Hence, the variation in detected intensity across the formed image can provide valuable insights into a material's local properties. 
The most common STEM measurement is annular dark-field imaging where the image intensity varies as approximately \(Z^{1.7}\), with Z being the atomic number. The generated images contain a wealth of information about the material structure and, if atomically-resolved, allow among other things, for mapping local polarization fields, identifying topological defects, and studying charge-density wave formation.
By using configurations with parallel detectors, the Z-contrast structure imaging can be combined with spectroscopic measurement – such as electron energy loss spectroscopy (EELS) or energy-dispersive X-ray spectroscopy - to probe materials' electronic functionality. Particularly, EELS can be used for probing the physics of collective excitations in nanoscale systems, as well as precise chemistry characterization. 
For example, STEM-EELS enables studies of local effects in plasmonic systems which are critical to the design of nanostructures with desired optical properties.
Finally, the recent advances in pixelated and multi-segmented detectors enable the acquisition of a diffraction pattern at each probe position, which constitutes the collection of techniques known as 4D-STEM. 
As a result, insights can be gained into the structure of electric and magnetic fields at the atomic scale, which in turn enable the study of chemistry of individual atomic defects, and mapping of interlayer spacing in quasi-2D materials (to name a few examples). 

\subsubsection{Microscopy Workflows}
\label{sec:microscopy_workflow}

A typical experimental study in STEM-EELS and 4D-STEM proceeds as follows. First, an annular dark-field scan over a relatively large field of view is acquired. Then, the regions for spectral or diffraction imaging, either single-point spectroscopy or a grid of points, are manually selected based on operator's intuition. The acquired data is usually stored at local resources or commercial cloud (e.g., Dropbox, Google Drive, etc.) and analyzed after the experiment is completed. The standard post-acquisition analysis of the hyper-spectral EELS data is performed via linear unmixing/decomposition techniques such as non-negative matrix factorization.
For 4D-STEM data, the processing is based on advanced analyses involving physics-based inversion. 
Recently, authors in~\cite{AE_STEM_EELS,AE_STEM_4D} demonstrated an approach for an 'intelligent' probing of dissimilar structural elements to discover in the automated fashion a desired physical functionality in STEM-EELS and 4D-STEM experiments. This approach utilizes deep kernel learning (DKL) to inform the next measurement by continuously learning a relationship between the local structure visualized via the dark-field STEM image and EEL spectra or 4D-STEM diffraction patterns. However, it is currently limited to relatively small data volumes as the DKL model training and inference are performed using local or on-board computational resources.
The remote computations and steering of the microscope based on analyses of measurements, by automated codes or manual operations, will contribute to the effectiveness of these workflows, and  eSolutions that enable them are our main goal of this paper.

\subsubsection{Nion Swift Software}

Here we focus on the Nion Swift software, which is open source, can be run on multiple OS platforms, and provides access to almost every aspect of the microscope. Note that other software platforms exist for different electron microscopes. 
The STEM microscopes considered here are controlled via Nion Swift software that provides a Graphical User Interface (GUI) and a python-based API.
It is installed and executed inside a Python virtual environment on the control computer, which is typically a Windows workstation co-located with the instrument. 
Its API provides instrument commands executed in a python console for measurement collection, microscope positioning and other tasks~\cite{swiftapi}.
A STEM instrument simulator, called \textit{nionswift-tool}~\cite{usim}, is also provided as an open source software package that provides an off-line Swift environment
identical to the physical installation.

\subsection{ORNL STEM and Servers}
\label{subsec:Federation Design}
\begin{figure}[t]
\centering
\includegraphics[width=0.47\textwidth]{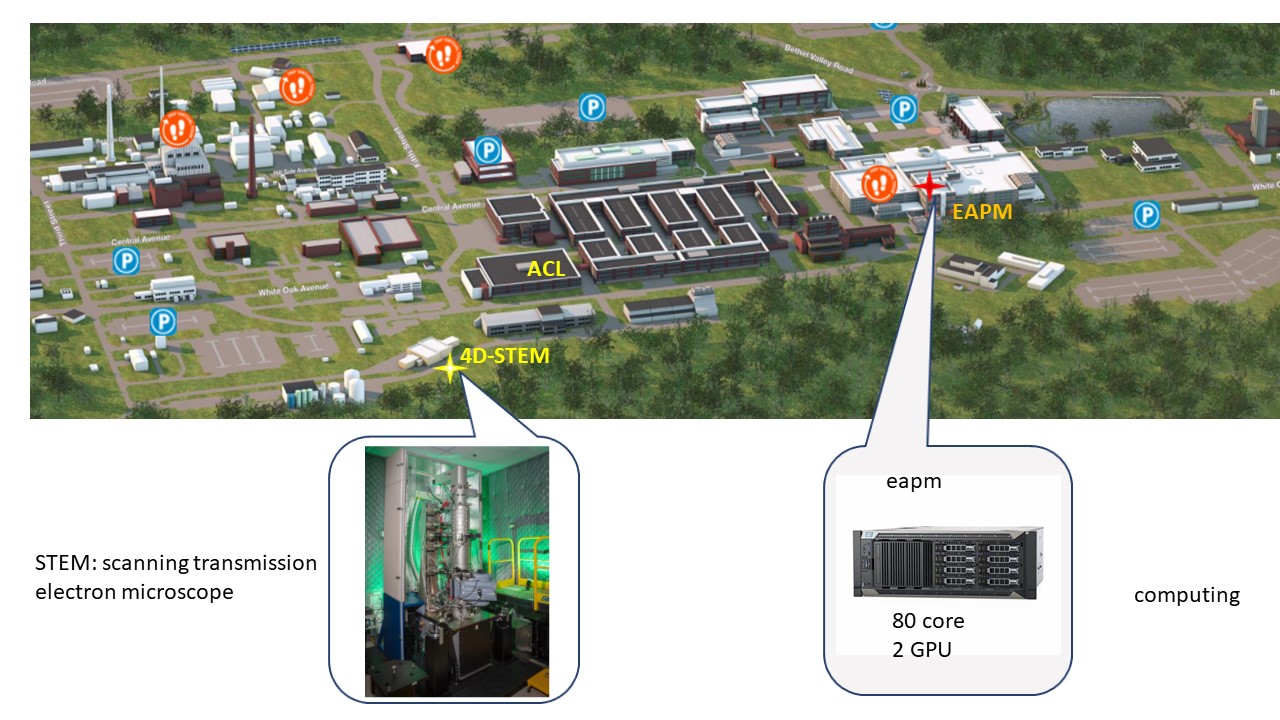}
\caption{\small The STEMs and computing workstations are in two ORNL facilities at separate physical sites and are connected to different site networks separated by firewalls. They include U100 and U200 Nion Microscopes and NAS systems at AML of CNMS, and eapm computing system with 80 CPU cores and two GPUs at K200 facility.} 
\label{fig:u100_eapm}
\end{figure}


The Nion STEM microscopes U100 and U200 of ORNL's Center for Nanophase Materials Science (CNMS) are located at
the Advanced Microscopy Laboratory (AML) site. 
The computing systems, including a server with 80 CPU cores and 2 GPUs, called {\bf eapm}, are located in a data center facility at a different site,
as shown in Figure~\ref{fig:u100_eapm}.
These facilities are serviced by different site networks which are separated by firewalls. 
A Nion microscope with an attached AS2 controller is controlled by the instrument control computer running the Swift software; they are connected over a local hub network which is not routed to other site networks.
The scientists conduct the microscopy experiments typically using the Swift GUI accessible via the control computer which is dual-homed to connect to a hub and site networks under strict firewall configurations.
Once the measurements are collected, they may be transferred to a NAS system connected to the microscopes and the control computer over the local hub network. In current workflows, the measurement collection and positioning commands are manually executed using Swift software, and data may be directly transferred to a local NAS. 

\subsection{Related Microscope and Instrument Control System}

SerialEM~\cite{mastronarde2005automated} is an electron microscope controller software enriched with complement applications for image acquisition, (pre)processing, display, buffering and file saving. It provides tools for supporting the essential microscopy operations, like tilt series, 3D recon­struc­tion and single-particle reconstructions. SerialEM is equipped with GUI, Python APIs as well as built-in script commands to support the data acquisition and processing. The controller supports a wide range of electron microscopes and Complementary Metal Oxide Semiconductor/Charge Coupled Device (CMOS/CCD)-based cameras. 
The Nion microscopes considered in this paper are not supported by SerialEM, but our Pyro-based solutions are in principle implementable using Python APIs. 

EPICS~\cite{epics} is an open-source toolkit used for distributed control of scientific instruments at experimental facilities, for example, Spallation Neutron Source (SNS) at ORNL and Advanced Photon Source (APS) at Argonne National Laboratory (ANL). It supports application interfaces and networking protocols for hardware components of instruments, such as sensors, motors, detectors, and magnets. It provides interfaces to Input/Output Controllers (IOCs) and Process Variables (PVs) using command lines (e. g., caget and caput to read and write PV values) or graphical interfaces. It also provides networking and interfaces access, known by Channel Access (CA) and Client CA (CAC), allowing science users to access, collect measurements and supply configuration parameters for targets used in science workflows.

Tango Controls~\cite{tango} is an open-source framework that manages a variety of systems and hardware types. It is applicable to different systems, including Distributed Control Systems (DCS), Integrated Control Systems (ICS)  and Supervisory Control And Data Acquisition (SCADA) systems, for instance, Machine to Machine (M2M), Internet of Things (IoT), Industrial  IoT (IIoT) applications, as well as national experimental facilities like synchrotrons. Tango provides classes for different instrument hardware, called the hardware classes, and it is scalable to build and integrate new hardware classes
. Tango also supports a range of programming languages, such as C++, Java and Python, for developing, controlling, and building hardware classes and framework software modules. Furthermore, various deployment modes are implemented with Tango for achieving autonomous and scalable control, for example, in-situ and distributed deployments, as well as remote steering functionalities via client-server or web client deployments.
The Nion microscopes at ORNL are not supported by EPICS or Tango controls, and consequently our solutions are primarily designed for their Swift software.


\section{STEM Ecosystem Design and Implementation}
\label{sec:design_and_implementation}

The implementation of a STEM ecosystem over an infrastructure, such as ORNL, requires the development, integration, and/or implementation of network and system configurations as well as new and available software modules.

\subsection{System Challenges and Solution Approach}
\label{subsec:Challenges and Solution Approach}

\begin{figure}[t]
\centering
\includegraphics[width=0.5\textwidth]{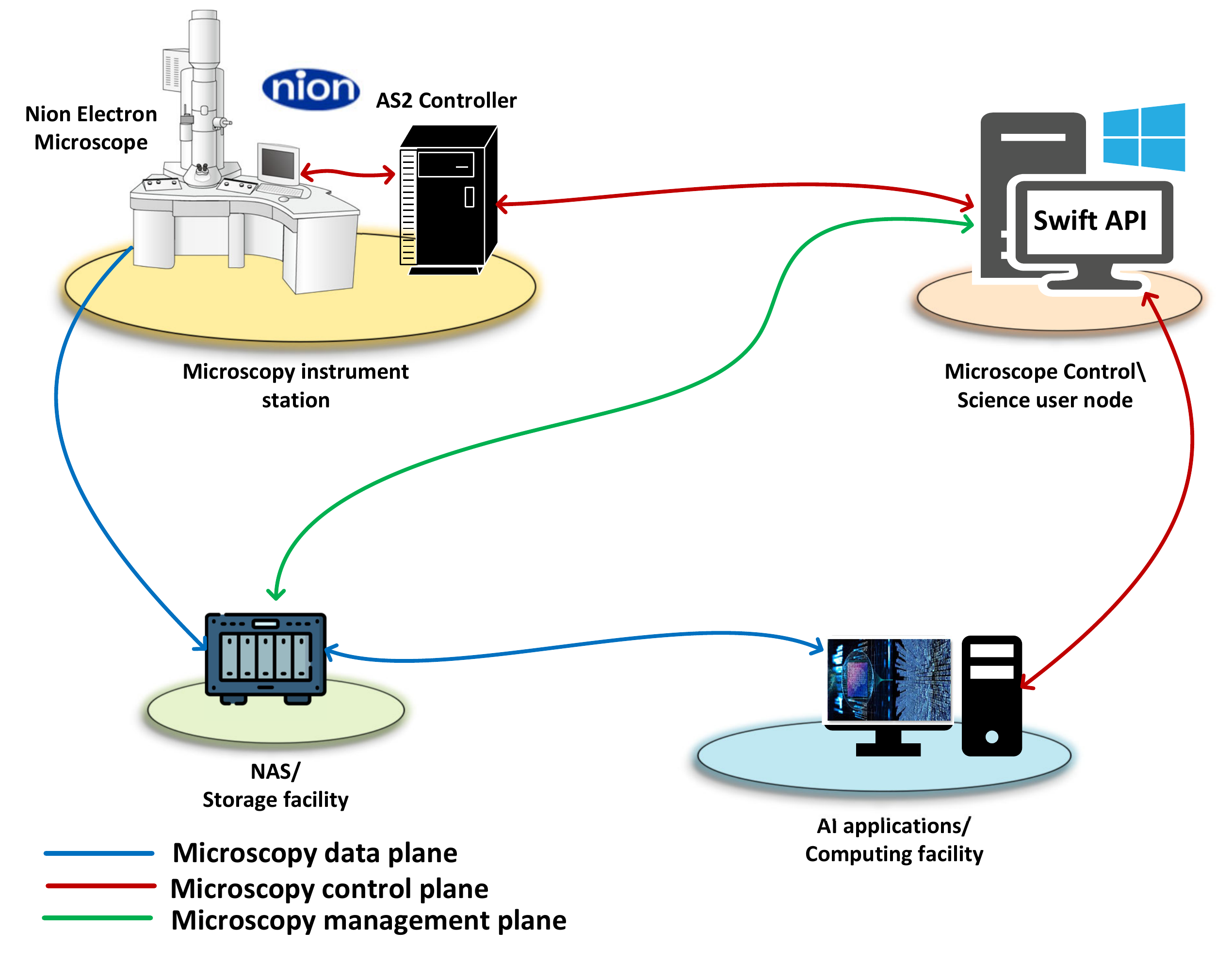}
\caption{Separate end-to-end control and data channels utilize Pyro client-servers and NAS, respectively, to support microscope steering and measurements collection needed for remote, automated experiments.} 
\label{fig:microscopy_workflow_integration}
\end{figure}

A STEM ecosystem needed to support remote autonomous experiments must address the following challenges:
\begin{itemize}

\item \textit{Local and remote access:} Microscopes are manually operated using custom software installed on control computers co-located with instruments, which are typically connected only to local hub networks. The access is needed to both their hardware and software over networks that are protected by access controls and firewalls.

\item \textit{OS and software:} Microscope software is typically proprietary and runs on Windows OS, and the measurements are stored on the local computer, and the storage systems often use custom mechanisms and formats. Both control and data access from remote Linux servers may be required, which entails interfacing different OSs, including programming environments, file and data formats, and host firewalls and access mechanisms.

\item \textit{Networking:} Network connections to control computers carry both measurements and control traffic, typically, over the same IP path and network interfaces. Over long network connections, this non-separation can potentially lead to the loss of instrument control when large measurement transfers occupy the entire available bandwidth. Suitable end-to-end network channels and mechanisms are needed between the microscope control computers and remote servers.
\end{itemize}

We propose a design based on separate end-to-end channels for control and data that are serviced by software modules that communicate across Windows and Linux OS. The control channels enable the scientists and automated codes to remotely access the microscope control computer and execute steering commands (Section III-B). The microscope measurements are collected on the NAS and made available on remote computing systems. 
Figure~\ref{fig:microscopy_workflow_integration} shows this design for ORNL ecosystem for the infrastructure described in previous section. 
We configure the NAS to export its file system (Section III-C), thereby making it available for analyses codes that utilize  powerful remote computing systems such as servers with multiple GPUs.

\begin{figure*}[thb]
\centering
\includegraphics[width=1.0\textwidth]{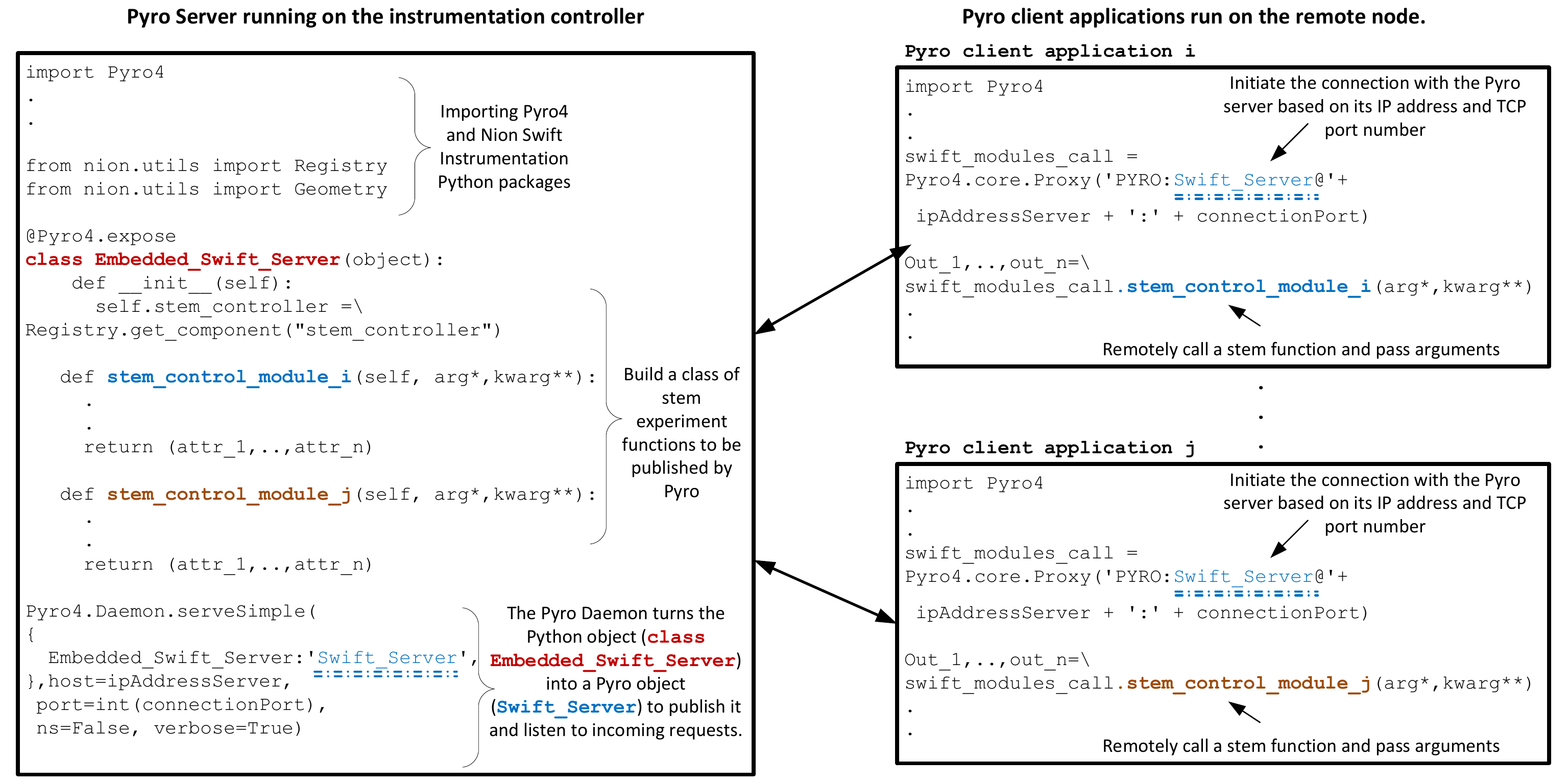}
\caption{Pyro client-server communication applied between the instrument control node and a compute system.} 
\label{fig:Pyro_server_client_setup}
\end{figure*}


\subsection{Control Channel}
\label{subsec:Control Connections}

A control channel is used to remotely access the instrument control computer for sending control commands and parameters to steer the microscope experiments. Upon command execution, the control computer may send back the results or other data. We developed Pyro client-server codes to support remote steering of microscope experiments across the ecosystem. Pyro provides a Python API for network access~\cite{pyro}, and we installed it on the control computer's Swift virtual environment and on the remote computing systems.
Figure \ref{fig:Pyro_server_client_setup} illustrates the developed solution of Pyro client-server communication across the ecosystem. 

Our Pyro server embeds python objects corresponding to STEM experiment tasks and makes them accessible and executable over control channels that span the ecosystem. These task codes are developed by us using Swift instrument commands~\cite{swiftapi} by utilizing APIs executable only under the control computer's Swift python environment. As shown in Figure \ref{fig:Pyro_server_client_setup}, the developed python object \textit{Embedded\_Swift\_Server} is a python class with multiple functions that encapsulate the STEM tasks. The Pyro daemon turns the python object into a Pyro object (\textit{Swift\_Server}) and publishes it via the control node's IP address and a TCP port to be accessible over the control channel. The Pyro server application is called from the \textit{File$\xrightarrow{}$Scripts} option in Swift GUI, and initiates as a background process which listens for incoming requests to execute microscope tasks using API commands.

Pyro client applications can be executed concurrently on multiple remote computing systems across the ecosystem. The client applications communicate with the Pyro server to execute the exposed functions on the control computer, as shown in Figure~\ref{fig:Pyro_server_client_setup}. A client initiates a connection with the server using a {Pyro object URI}, which is a resource identifier in the format: \textbf{PYRO:objectid@IP:TCP port}. The \textbf{objectid} refers to the Pyro object published by Pyro daemon running on the control computer, which is \textit{swift\_Server} in our implementation shown in Figure~\ref{fig:Pyro_server_client_setup}, and  \textbf{IP} is the network address of the control computer. The \textbf{TCP port} specifies the communication port between the Pyro client and server applications that allows them to communicate and exchange control messages. In our implementation, we developed a client application for each STEM experiment task exposed by the Pyro server. The client applications are called from python console or embedded in automated scripts, for example, using Jupyter notebooks, by passing the IP address of the control node and  control commands and parameters required to steer the microscopy experiments.
Overall, the approach of wrapping the microscope Swift APIs using Pyro client-servers enables this solution to scale and be portable across multiple STEMs and computing systems across the ecosystem.


We developed several microscope task modules, namely \textit{scan\_status} to get the current scan status, \textit{scan\_channel} to obtain measurements from a particular microscope channel, and \textit{probe\_position} to position the beam at specified coordinates. These modules are implemented as functions of the Pyro server (details in Appendix A). They are paired with the corresponding Pyro client modules,  \textit{check\_scan.py}, \textit{scan\_channel.py}, and \textit{probe\_position.py}, respectively. 
These client modules pass the user or machine-driven control commands to the Pyro server to execute microscope tasks.

The concept and codes for Pyro servers and clients are developed without requiring access to the physical infrastructure by using the VIT of ORNL ecosystem  with Nion Swift simulator, as described in Section \ref{sec:Virtual Infrastructure Twin}. These steps took several days, and once matured, the codes were tested and demonstrated over ORNL STEM ecosystem which required the physical access and an operator for the microscope to ensure safety, as described in Section \ref{sec:experimental_setup}.

\subsection{Data Channel}
\label{subsec:Data Connections}

The data channel makes the measurements available at the NAS connected to the control computer and also at the remote servers of the ecosystem, up on the execution of commands either locally or remotely over the control channel.
The Swift software is configured to store the measurements on the Windows-based NAS as files.
We implement the data channel by remote mounting the NAS data files using Samba and Common Internet File System (CIFS) file sharing, which provides access across different operating systems, in particular, Linux servers. The Windows-based NAS files are natively available on the Linux-based servers to perform scientific analysis and computations. The access privileges to the computing nodes and  Samba/CIFS file sharing are configured to allow the users to access the NAS files across the ecosystem.
This access via one-time setup makes it persistent across the ecosystem.
Indeed, the cross-facility mounting of microscopy data automates large-scale data transfer and makes it available for the computations across the ecosystem. File transfer tools such as GridFTP \cite{gridftp} or Globus \cite{globus} applications require additional hardware and/or software, including licenses and credentials, which may become difficult to manage by the scientists. 
The Pyro communication is not used for large measurement transfers since they are limited by memory size and they can also generate cross traffic that can potentially limit the control traffic. 

The data channel transparently transfers high-volumes of microscopy measurements over network connections that are separate from control channel connections, thereby  mitigating the impact of large measurement transfers on the steering operations, as described in next section.



\begin{figure}[t]
\centering
\includegraphics[width=0.45\textwidth]{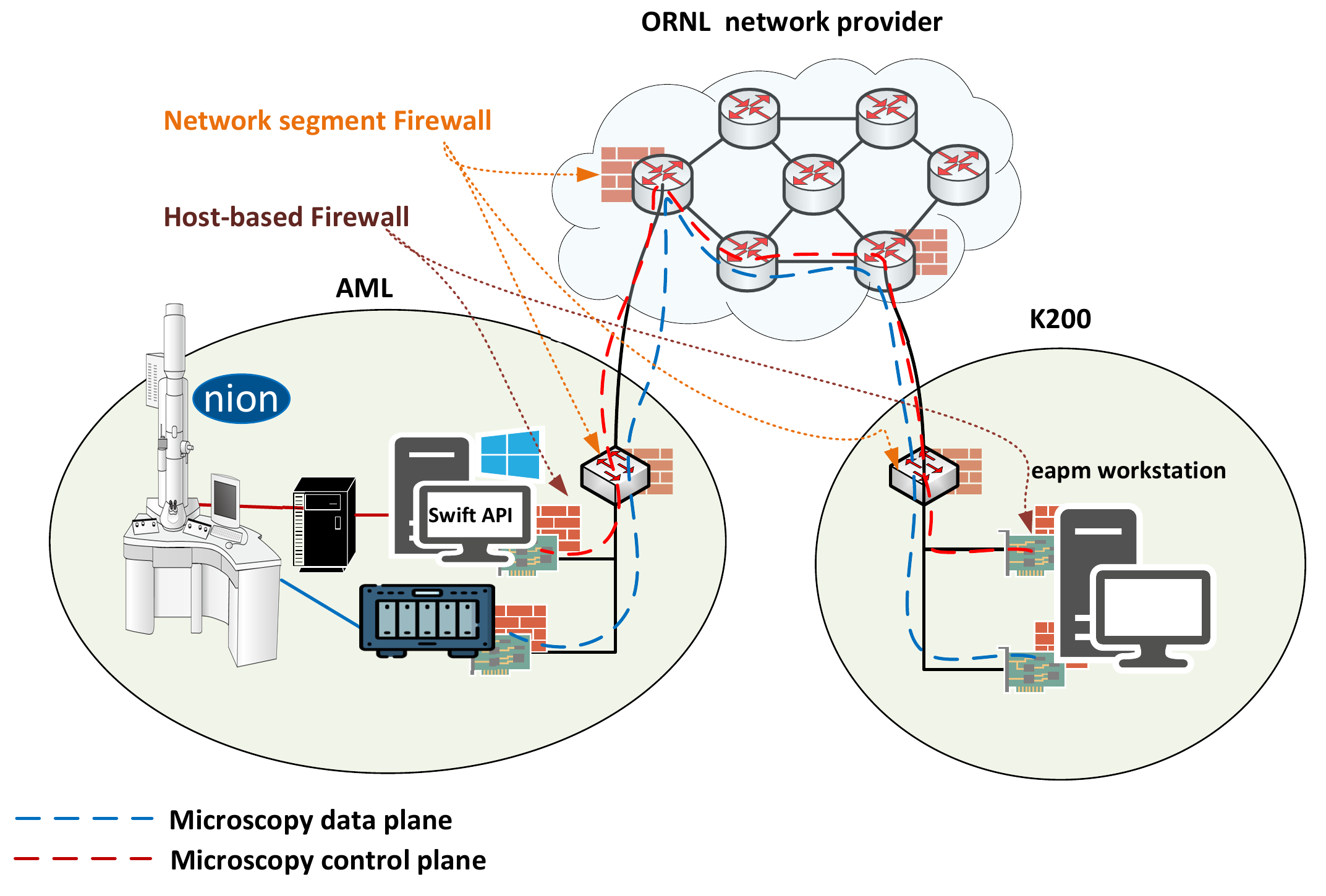}
\caption{Control and data channels are enabled by configuring the network and host firewalls, and are separated by two NICs on computing server.} 
\label{fig:control-data-channels}
\end{figure}

\subsection{Networks, Access and Firewalls}

The microscope computers and remote servers are located at different networks separated by network firewalls and Windows and Linux host firewalls, as shown in Figure \ref{fig:control-data-channels}.
Both the data and control channels are enabled by inserting various firewall rules to support file mounting between the NAS and computing servers over the data channel, and open communication ports for Pyro servers and clients over the control channel.
Their rules are inserted both at the firewalls separating these two networks, and also on the Windows and Linux hosts.
Two physical interfaces are configured on the compute servers with different IP addresses to separate the control channel traffic between Pyro server and clients, and remote mounting the NAS CIFS file system.
The end points of both control and data channels are two separate NICs on the same compute servers but their other ends points are on separate systems, namely, the control computer and NAS.


\section{experimental setup and Demonstration}
\label{sec:experimental_setup}

\begin{figure}[t]
\centering

\includegraphics[width=0.45\textwidth]{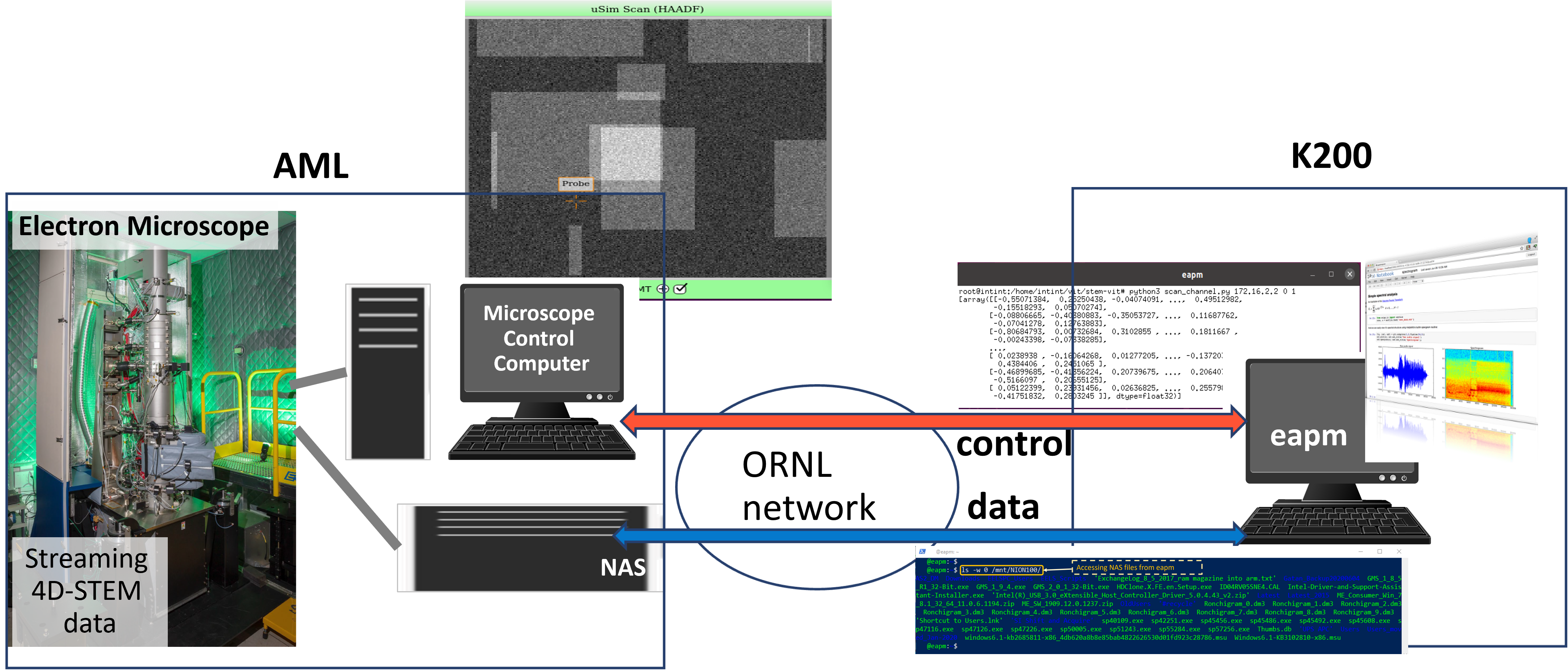}
\caption{Data and control channels between STEMs at AML and computing servers in K200 in ORNL  infrastructure.} 
\label{fig:Remote steering Microscopy experiments}
\end{figure}

The ecosystem capabilities described in the previous section are implemented on U100 and U200 microscopes at the AML science facility, and on the eapm workstation at the K200 computing facility, which are both parts of ORNL physical infrastructure. 
The experimental setup shown in  Figure~\ref{fig:Remote steering Microscopy experiments} is used for remote steering and measurement transfer operations carried out  between the microscope and its NAS system and on the eapm at K200. 


\subsection{Steering experiments over ORNL ecosystem}
\label{subsec:Steering experiments over ORNL ecosystem}
\begin{figure*}[t]
	\centering
  \includegraphics[width=0.8\linewidth, height=22em]{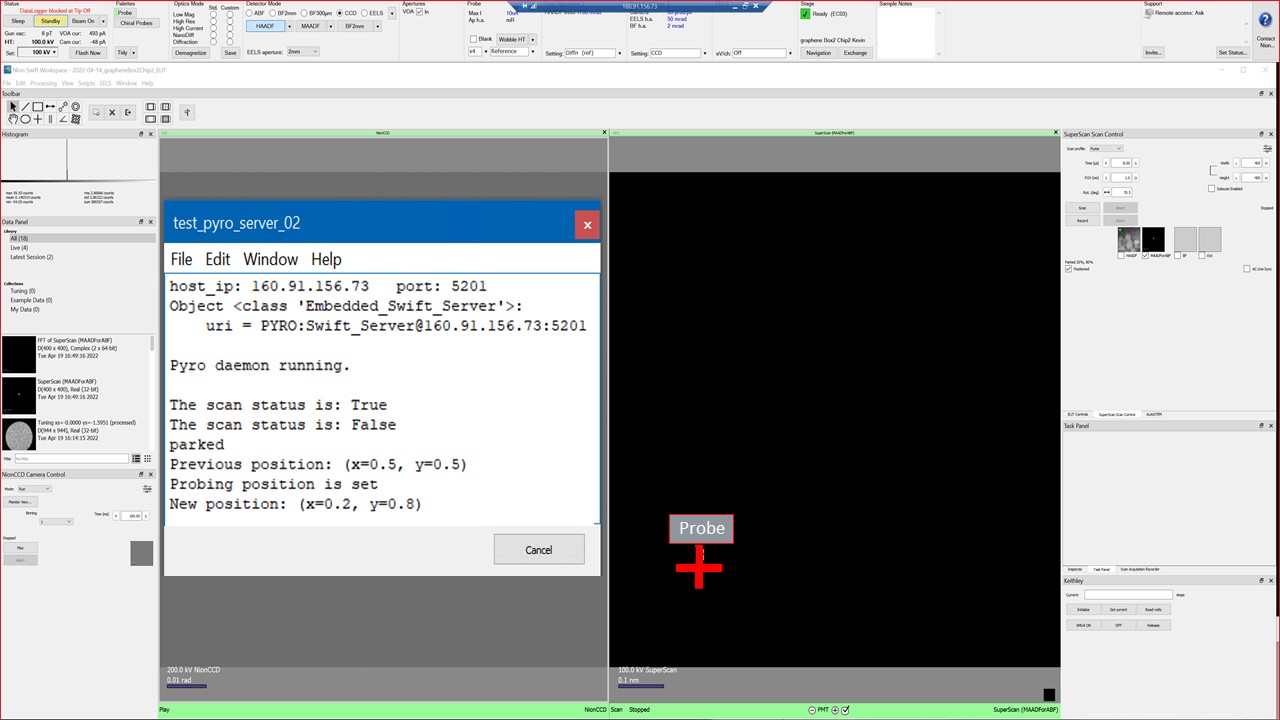}
\caption{Screenshot of Swift GUI running at U200 microscope control compute at AML, ORNL. The Pyro server output corresponds to the execution of Pyro client commands from eapm computing system located at K200 and the resultant positioning of microscope probe.}
\label{fig:U200_screenshot}
\end{figure*}

We have successfully tested the microscopy control channel between the {U200} microscope  and {eapm} computing system by steering the microscope experiments over the physical infrastructure of ORNL ecosystem (Figure \ref{fig:u100_eapm}). 


We integrated a number of Python functions for STEM APIs to obtain the beam status and other microscope parameters as well as to position the beam.
In particular, the functions \textit{scan\_status} and \textit{probe\_position} (explained in Appendix \ref{subsec:scan status} and \ref{subsec:prob_position}) are the corresponding Pyro server objects on the {U200}  control computer. The microscope is steered by \textit{check\_scan.py} and \textit{probe\_position.py} Pyro client applications running on {eapm}. 

On the {U200} control computer, the response to cross-facility communication with {eapm} system is shown in the screen shot in Figure~\ref{fig:U200_screenshot}. 
The Pyro server's response to the scan\_status commands from Pyro client on eapm is shown in the console, as
two True and False responses which are also sent back to eapm.
The response to probe\_position command is shown in console as the previous probe position ($x=0.5$ and $y=0.5$) and  the new probing position at $x=0.2$ and $y=0.8$ from Pyro client. The resultant new probe position is depicted on the right side window of Figure~\ref{fig:U200_screenshot}.

\begin{figure}[t]
\centering
\includegraphics[width=0.5\textwidth]{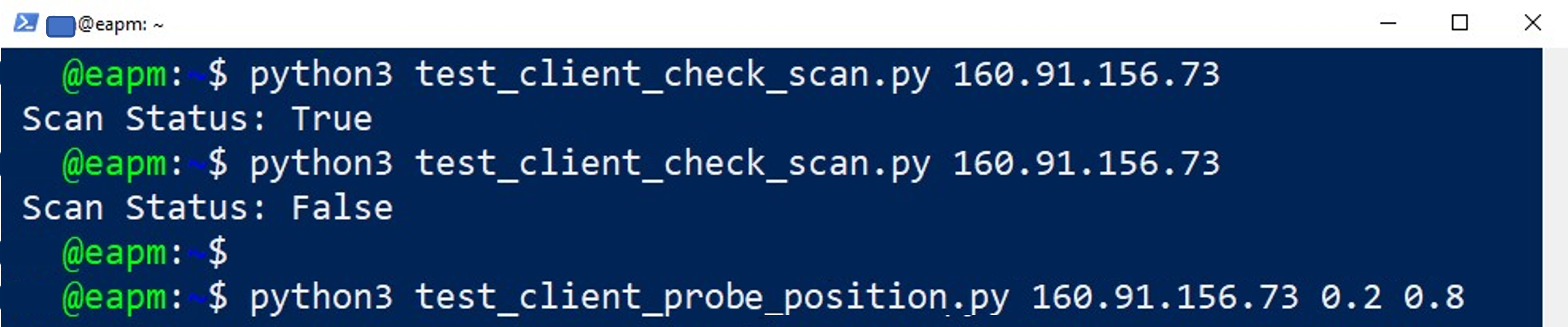}
\caption{Output of Pyro client on eapm in checking scan status and probing position on U200 over ORNL physical infrastructure.} 
\label{fig:ornl_k200_eapm}
\end{figure}

On the eapm computing system, the output of Pyro client applications for executing remote microscope tasks is shown in
Figure~\ref{fig:ornl_k200_eapm}. The scan status of the U200 microscope, denoted by the IP address (160.91.156.73), is frequently checked using \textit{check\_scan.py} application. The scan status is {True} while running a physical scan on the control node via Swift GUI,  and when the scan is completed its status becomes {False}. Also, a new scan position is sent to the microscope using \textit{probe\_position.py} application with coordinates ($x=0.2$ and $y=0.8$) to which the Pyro server responded by positioning the microscope.

These Pyro codes are initially developed without requiring physical access using VIT as described in Section \ref{sec:Virtual Infrastructure Twin}, and are readily executed on ORNL infrastructure.

\subsection{Automated Data Transfers}
\label{subsec:Automated Data Transfers}

\begin{figure}[t]
\centering
\includegraphics[width=0.47\textwidth]{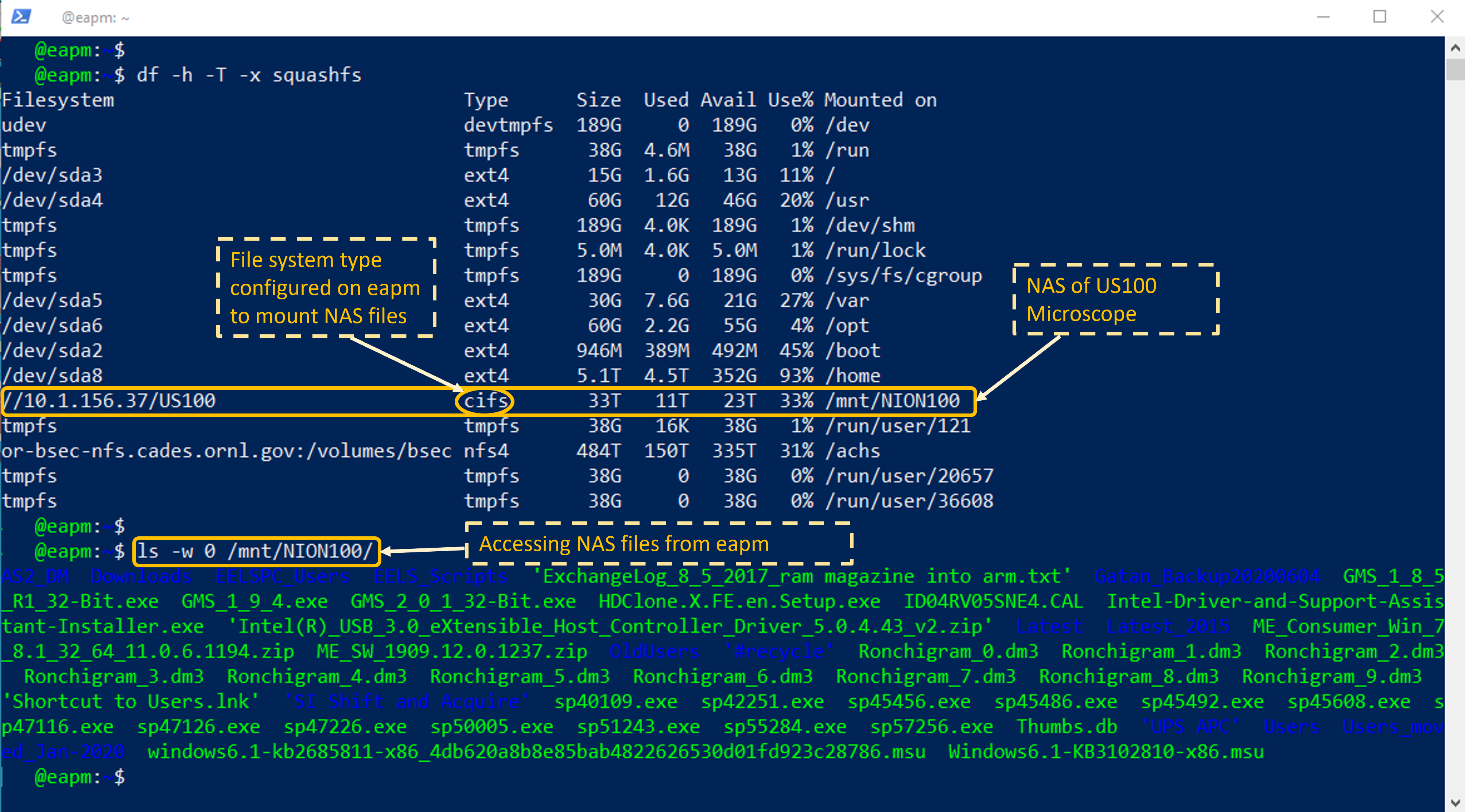}
\caption{NAS mounted on eapm for automatic data transfer.} 
\label{fig:eapm_nas}
\end{figure}

We deployed and tested the proposed data channel  over ORNL physical infrastructure by remote mounting the microscope measurements directory of NAS devices on eapm. 
For example, the US100 NAS system, accessed by IP address 10.1.156.37, is mounted to eapm using CIFS file system. The authorized microscopists and automated codes seamlessly access NAS files at the mounted directory \textbf{/mnt/NION100} and utilize them in computations on the computing system.
A screenshot in Figure~\ref{fig:eapm_nas} shows these U100 measurement files on NAS being available on eapm. 

    \section{STEM Virtual Infrastructure Twins}
\label{sec:Virtual Infrastructure Twin}

The development and implementation of a STEM ecosystem requires various software components and network and system configurations to be designed and tested.
These workflows that utilize networked computations and instruments present challenges that are not typically faced in (pure) computing ecosystems. 
It is too expensive and potentially disruptive for the whole STEM ecosystem to be available during the entire development and testing period, particularly, in early stages. 
More generally, the ecosystems, such as a multi-site STEM complex, may not be available while they are being designed and developed, and indeed, may not always be needed, for example, instrument time is not usually necessary to debug network code.
In this context, we propose employing a STEM VIT (S-VIT) as an enabling software tool to be used prior to the field deployment.
In particular, ORNL S-VIT (OS-VIT) is used to develop the Pyro steering codes described in previous sections, and an additional Multi-site S-VIT (MS-VIT) is developed in this section to show their applicability to ecosystems that span multiple sites connected over wide-area networks.

\subsection{VIT: Concept and Design}
\label{subsec:VIT: Concept and Design}

VIT emulates the network and computing components of the ecosystem, and incorporates instrument software simulators, such as Nion Swift simulator in S-VIT.
It provides a software environment nearly identical to the physical ecosystem to support early and continual development, testing and design space explorations. It is implemented using mininet~\cite{mininet} by using virtual hosts to execute and test scientific applications (including microscopy applications and simulations), and to emulate the network infrastructure using virtual switches and routers. The emulated components communicate over virtual links that reflect the physical network infrastructure links. VIT is packaged as a portable virtual machine to facilitate the development process among scientific communities. Once the VIT-based solutions are suitably tested, they are ready to be field-deployed and integrated into the physical infrastructure. 
Previous VIT implementations address other scenarios  including software defined networking \cite{Liuetal2018aiscience}, EPICS~\cite{osti_1817424} and federation software stack \cite{alnajjaretal2020netsys}.

\begin{figure}[t]
\centering
\includegraphics[width=0.5\textwidth]{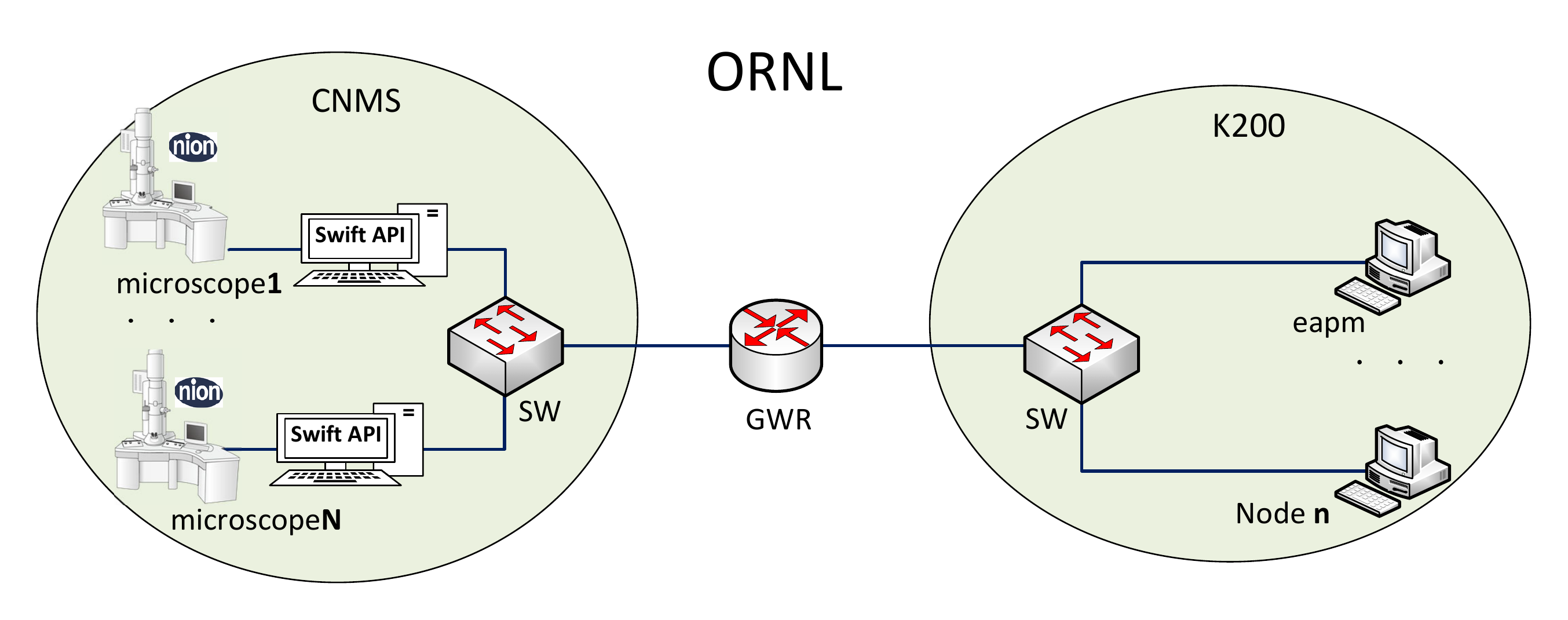}
\caption{OS-VIT: Emulation of ecosystem of ORNL STEM infrastructure.} 
\label{fig:vit_microscope}
\end{figure}

\subsection{ORNL STEM Ecosystem: OS-VIT}

The OS-VIT for developing and testing STEM workflows across the ORNL infrastructure is shown in Figure~\ref{fig:vit_microscope}. It consists of two facilities,  K200 computing facility and CNMS microscope facility. It subsumes the physical infrastructure used in experiments in Section \ref{sec:experimental_setup}, and  provides additional systems.
The K200 facility emulation provides  multiple virtual hosts (including one for eapm) that are used for computations and remote access. 
The simulation of the CNMS facility includes multiple microscope control computers (related to U100 and U200 at AML) that support simulated STEM experiments. 
The control computers are emulated using virtual hosts that run \textit{nionswift-tool} simulator that includes Swift GUI.
The facilities' devices are connected via virtual switches, which in turn, connect the two facilities via a Gateway Router (GWR).

\label{subsec:VIT: Microscope Instrument Control}
\begin{figure}[t]
\centering
\includegraphics[width=0.47\textwidth]{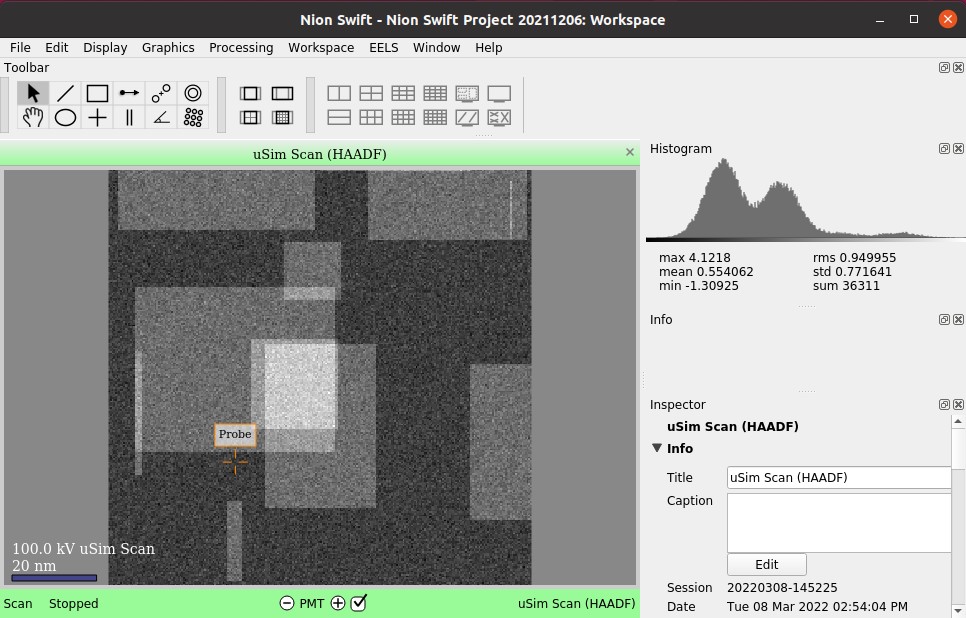}
\caption{Swift GUI runs on microscope1 node at CNMS.} 
\label{fig:vit_mscop1_nionswift2_microscopy_wf}
\end{figure}

\begin{figure}[t]
\centering
\includegraphics[width=0.47\textwidth]{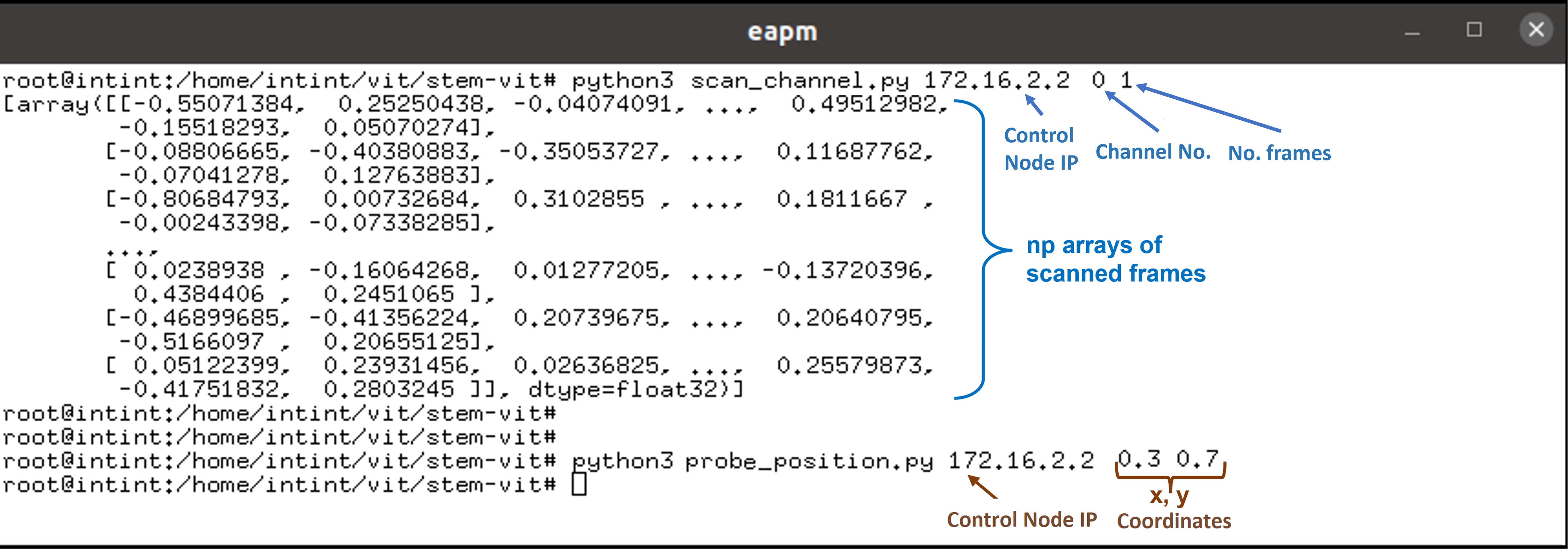}
\caption{Running Pyro client applications on eapm node at K200. The applications represent scanning a microscope channel and probe certain position after the scan is complete.} 
\label{fig:vit_eapm_microscopy_wf}
\end{figure}

\begin{figure}[t]
\centering
\includegraphics[width=0.47\textwidth]{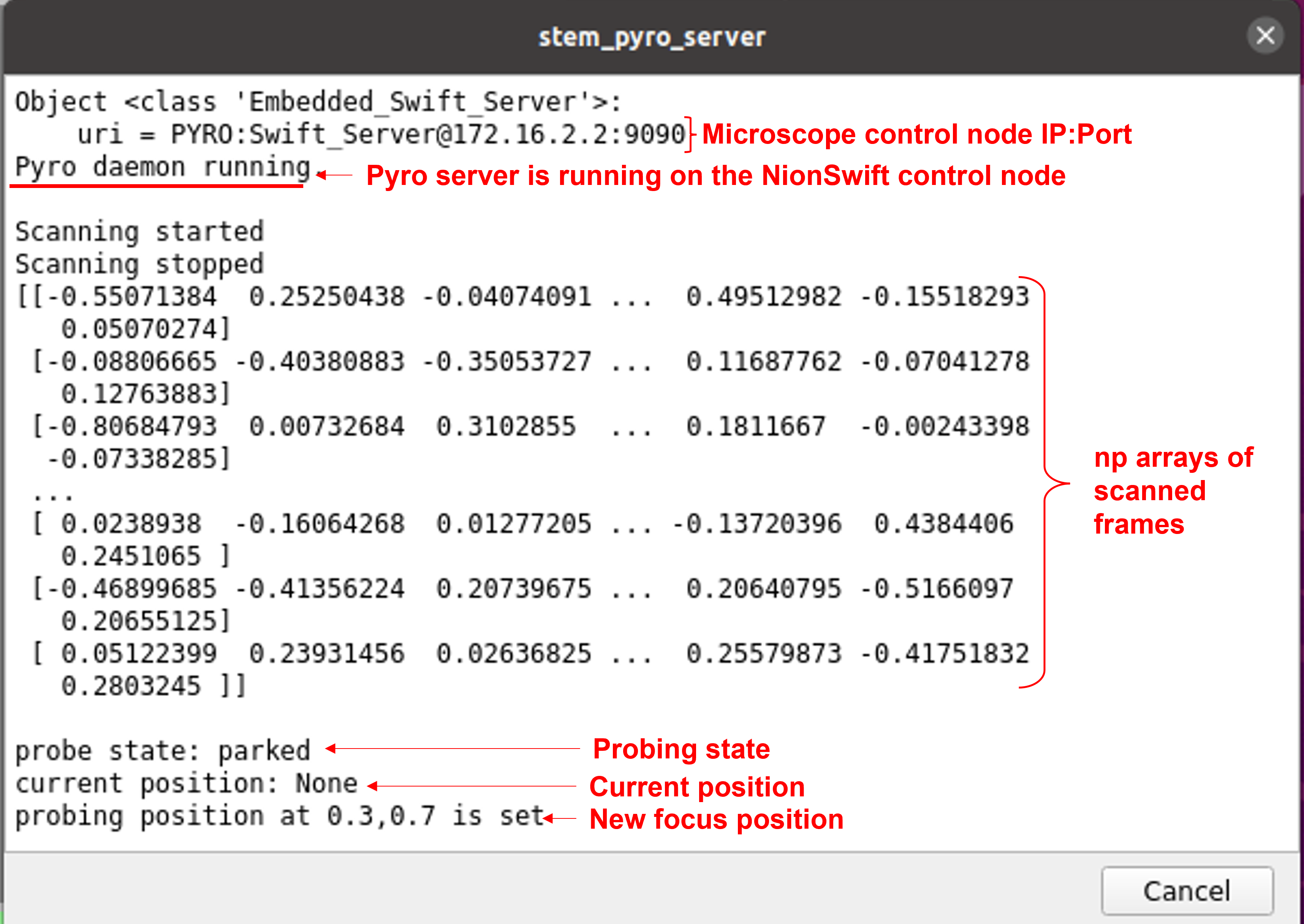}
\caption{Script window at nionswift-tools runs at microscope1/CNMS shows  running  Pyro  server exposing  STEM experiments across the emulated ORNL infrastructure and logs the status of the execution.} 
\label{fig:vit_mscop1_nionswift_pyro_microscopy_wf}
\end{figure}


The OS-VIT provides remote access from eapm at K200 to microscope1 control computer at CNMS to steer STEM experiments, send back the data for potential use in computations and change the microscope focus position (using \textit{scan\_channel} and \textit{probe\_position} Pyro STEM tasks). 
The workflow tasks described in Section \ref{sec:experimental_setup} are first carried out on OS-VIT.

First, the Swift API is run on the microscope1 node to load the Swift GUI, as shown in Figure~\ref{fig:vit_mscop1_nionswift2_microscopy_wf}. 
Next, the Pyro server module is loaded on Swift GUI and run as a daemon that waits for Pyro client communications across the ecosystem. 
Figure~\ref{fig:vit_eapm_microscopy_wf}  shows the output of  Pyro client applications on eapm:  \textit{scan\_channel.py} is executed to scan channel number zero and gather data related to one frame at the control node, which is sent as a NumPy array to eapm for computations.

Once the channel scan finishes, the microscopist would be able to reconstruct and analyze the scanned data, including for particular positions. For example, The microscopists at eapm can change the focus position and collect new measurements. We demonstrated changing the probe position via executing the \textit{probe\_position.py} client application with new coordinates $x=0.3$ and $y=0.7$ that are sent as parameters of \textit{probe\_position} to microscope1. The effect of changing the focus position is depicted on the updated image in Figure~\ref{fig:vit_mscop1_nionswift2_microscopy_wf}.  The STEM experiments' status is interactively shown on the Script window at Swift API shown in Figure~\ref{fig:vit_mscop1_nionswift_pyro_microscopy_wf}, including the generated data from executing channel\_scanning and probe\_position tasks.

\begin{figure*}[thb]
\centering
\includegraphics[trim=0cm 0cm 0cm 0cm,width=0.7\textwidth]{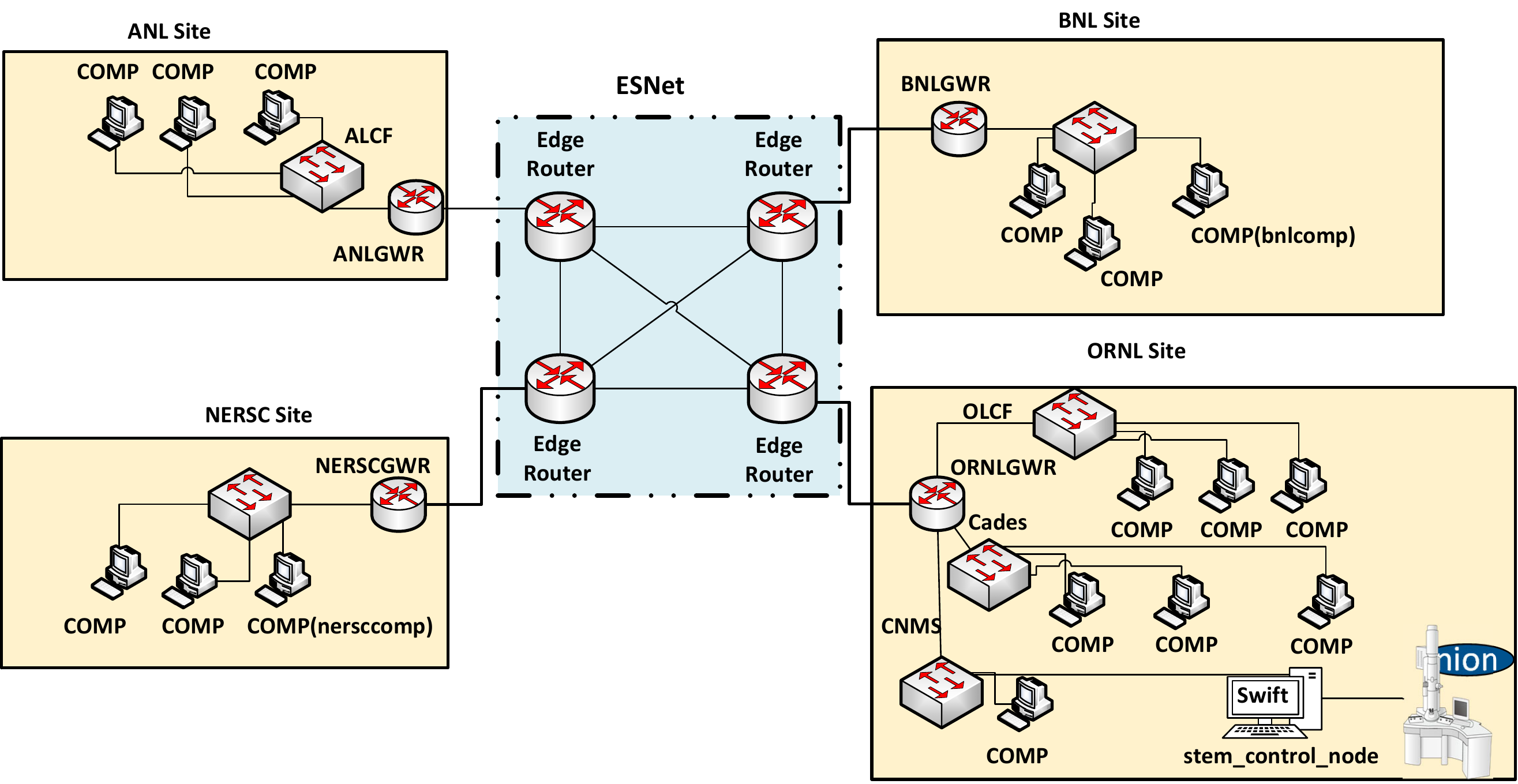}
\caption{MS-VIT: Emulation of STEM ecosystem of four DOE lab sites connected over a wide-area network.}  
\label{fig:4-sites_microscope_workflow}
\end{figure*}

\subsection{Multi-site STEM Ecosystem: MS-VIT}
\label{subsec:Concurrent STEM workflow control}

In addition to reflecting a specific infrastructure (as by OS-VIT),  VITs can be developed for more general purposes such as assessing new designs, building and testing more complex ecosystems that support more sophisticated scientific workflows.
In particular, we consider ecosystems of multiple sites,  each with several computing facilities and scientific instruments. 
MS-VIT is an emulation of an infrastructure of multiple Department of Energy (DOE) sites shown in
Figure~\ref{fig:4-sites_microscope_workflow}. This infrastructure consists of four DOE labs, namely ORNL, Brookhaven National Laboratory (BNL), Argonne National Laboratory (ANL), and The National Energy Research Scientific Computing Center (NERSC). These sites consist of computing systems and may  include scientific facilities, such as CNMS at ORNL. Virtual hosts (denoted by {COMP}) are incorporated as part of the ecosystem to execute scientific computations and STEM applications. For example, the Swift simulator (nionswift-tool) is run on a virtual host at CNMS representing the instrument control node that simulates Nion STEM experiments responses.
The ecosystem's site networks are connected to gateway routers, which are in turn, connected to edge routers of the wide-area network, namely, ESnet~\cite{ESnet}; the latter is emulated as a set of virtual routers with dedicated site-to-site connections.
Details of VITs used for multi-site federation implementation and a science use case using EPICS simulator, are described in~\cite{lcn2021-Al-Najjar-vfsie-paper} and \cite{lcn2021-Al-Najjar-vfsie-demo}, respectively.

We utilize MS-VIT to explore the feasibility of coordinating concurrent operations of an electron microscope by multiple STEM workflows across the multi-site ecosystem. In this scenario, we demonstrate having two microscopy workflows, namely {BNL-ORNL WF} and {NERSC-ORNL WF}, that are running concurrently. These workflows remotely access and execute the STEM experiments that are part of the Pyro object running on \textbf{stem\_control\_node} at CNMS/ORNL.
Such concurrent workflows are feasible independent of which (user's) sample is currently loaded because the Nion Microscope is equipped with a magazine system capable of carrying multiple samples that can be dynamically loaded during the experiments.
The {BNL-ORNL WF} remotely steers the microscope experiments from the \textbf{bnlcomp} compute node at the BNL while the other workflow steers the microscope from the \textbf{nersccomp} compute node at NERSC. The STEM tasks incorporated in this scenario are \textit{scan\_channel} and \textit{scan\_status}.

\begin{figure}[t]
\centering
\includegraphics[width=0.47\textwidth]{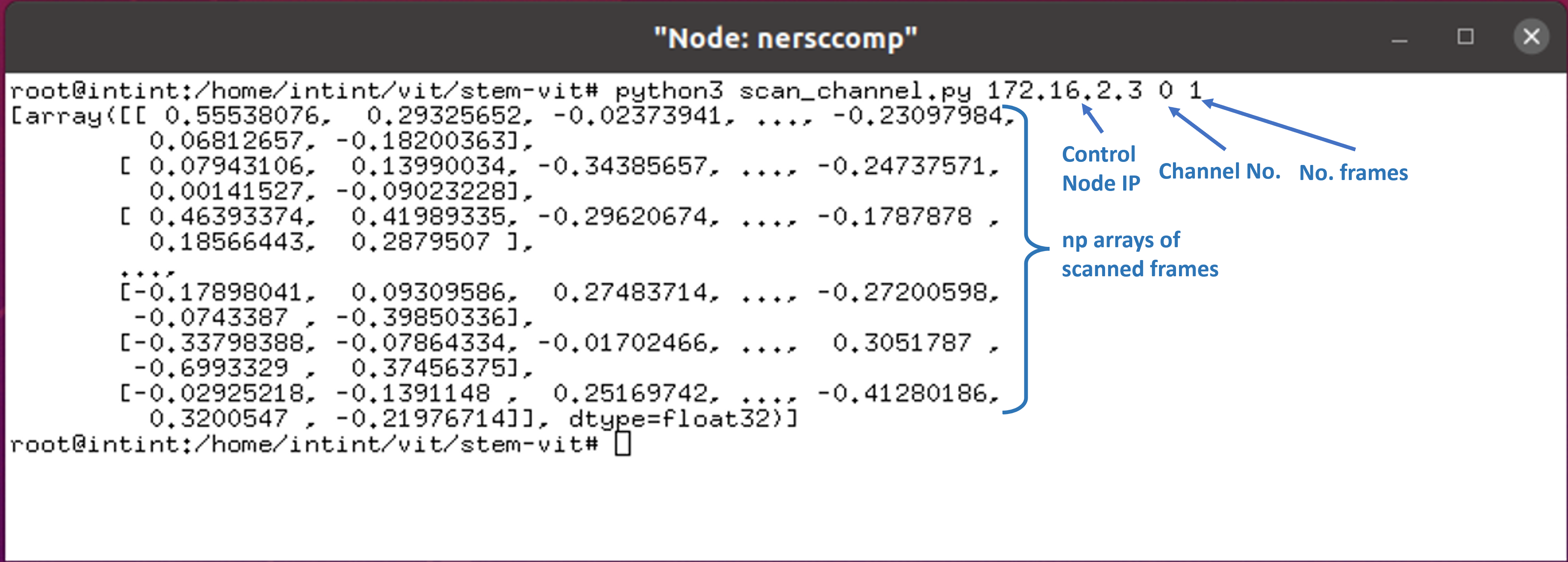}
\caption{Run a Pyro client module at a COMP node at NERSC site to scan a microscope channel.} 
\label{fig:vfsie_nersc_microscopy_wf}
\end{figure}

\begin{figure}[t]
\centering
\includegraphics[width=0.47\textwidth]{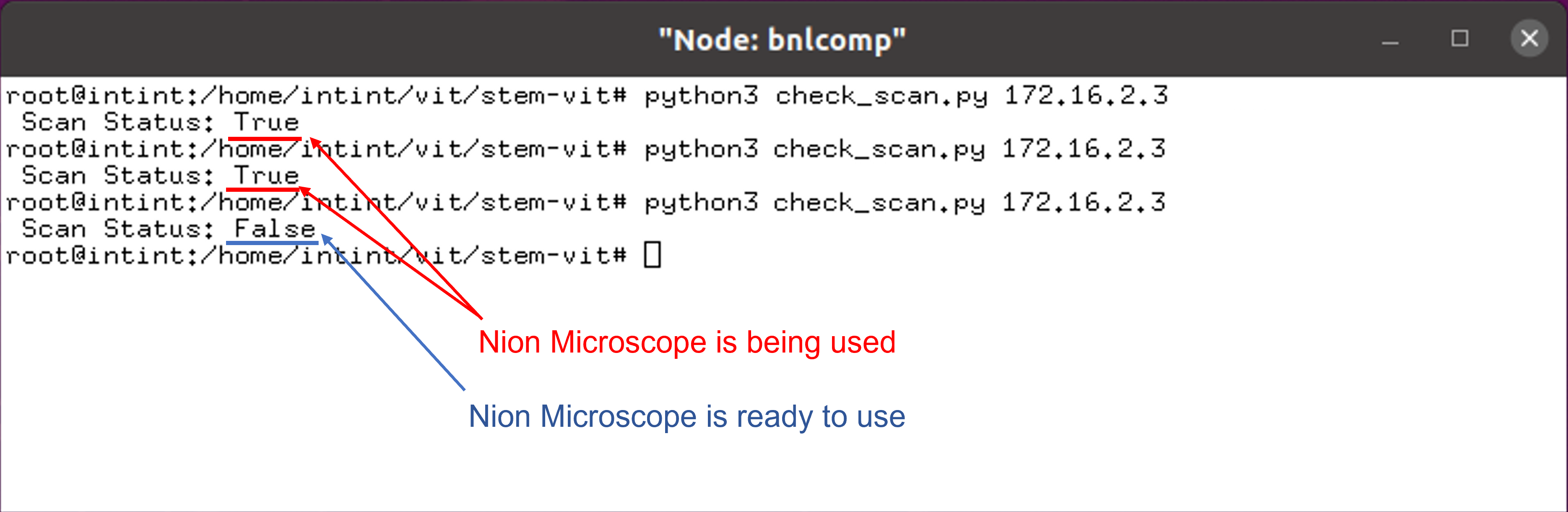}
\caption{Run a Pyro client module at a COMP node at BNL site to probe the scan status.} 
\label{fig:vfsie_bnl_microscopy_wf}
\end{figure}

Different science users initialize these workflows at NERSC and BNL sites and both communicate with \textbf{stem\_control\_node}. The scenario works as follows. A science user at the NERSC site initiates {NERSC-ORNL WF} that includes executing \textit{scan\_channel.py} on the \textbf{nersccomp} node, as shown Figure~\ref{fig:vfsie_nersc_microscopy_wf}. Concurrently, another science user at BNL is checking the scanning status of the Nion microscope, whether it is available to launch the {BNL-ORNL WF}. Figure~\ref{fig:vfsie_bnl_microscopy_wf} shows output of \textit{check\_scan.py} on the \textbf{bnlcomp} node. The results show \textbf{True} scanning status while {NERSC-ORNL WF} is running, and upon its completion scan status is \textbf{False} which is an indication of the availability of microscope at CNMS/ORNL. 

The MS-VIT reflects the current facilities, and can be extended to include those that are currently being  built or being designed for possible future deployments. Overall, the motivation for developing such multi-site ecosystem VITs range from developing solutions for certain current scientific instruments without requiring their physical access to future ecosystem designs. 

\section{conclusions and future work}
\label{sec: conclusion}

We presented eSolutions, both field-deployments and enabling virtual twins, to support the development and testing of remote experiment capabilities for science workflows over microscope ecosystems.
Together, they enabled the development of control and data channel implementations for a production STEM ecosystem, and also proof-of-principle demonstrations of a wide applicability of this VIT approach to multi-site microscope ecosystems.
The overall approach is applicable to other science scenarios, such as automated, remotely controlled chemistry and materials experiments.
In these cases, the custom software for instruments such as chromotographs, potentiostats, flow reactors, and others, may be leveraged to form ecosystems by wrapping their APIs using Pyro codes.

Future extensions of this work include GUIs that provide a comprehensive dashboard of the entire ecosystem for scientists and facility operators; performance assessment of the ecosystem including the data and control channels and workflows; and  the development of production quality software stacks for establishing and operating ecosystems by building up on our experimental codes.
It would be of future interest to develop similar solutions to other classes of microscopes by exposing their API or callable functions for network communications using Pyro modules, in particular, those supported by serialEM.
Other future work areas include integration of
machine-driven instrument control based on  AI/ML methods into microscope ecosystems, and frameworks that support integrated workflows, for example, Jupyter notebooks that integrate instrument operations and computations based on measurements.
It would of future interest to explore the applicability of the proposed solution  based on parallel data and control channels to other instruments such as those supported by EPICS and Tango controls.



\appendices

\section{Pyro Modules for STEM Tasks Using API}
\label{sec:Pyro Server: Microscope Commands}




This Appendix describes Nion microscope tasks developed as functions based on Swift API commands and exposed by Pyro server modules on the instrument control computer.

\subsection{Scan status}
\label{subsec:scan status}

This STEM task checks the scan status on the microscope. It is built as a function that returns \textit{True} to the client as a Boolean variable  if the microscope is occupied with scanning and \textit{False} otherwise. The function is listed below.

\begin{algorithm}[H]
\begin{algorithmic}[1]
\Statex \textbf{def} scan\_status(self):
\STATE \hskip1.0em scan = self.stem\_controller.scan\_controller
\STATE \hskip1.0em \textbf{return} scan.is\_playing
\end{algorithmic}
\end{algorithm}


\subsection{Scan\_channel}
\label{subsec:channel_scanning}

This STEM task function scans a microscope channel and transfers a number of frames of the data matrices corresponding to the scan. The function receives control parameters related to the channel number and number of frames from its peer Pyro client application, and returns arrays of the scanned frames in that channel. The code for scan\_channel function is shown below.

\begin{algorithm}[H]
  \begin{algorithmic}[1]
    \Statex \textbf{def} scan\_channel(self, ch, num\_frames):
    \STATE \hskip1.0em ct=1 \char"0023 constant time
    \STATE \hskip1.0em scan = self.stem\_controller.scan\_controller
    \STATE \hskip1.0em scan.set\_enabled\_channels([ch])
    \STATE \hskip1.0em frame\_parameters =$\backslash$
    \Statex \hskip1.75em scan.get\_current\_frame\_parameters()
    \STATE \hskip1.0em frame\_time =$\backslash$
    \Statex \hskip1.75em scan.calculate\_frame\_time(frame\_parameters)
    \STATE \hskip1.0em scan.start\_playing(frame\_parameters)
    \STATE \hskip1.0em time.sleep(frame\_time * num\_frames + ct)
    \STATE \hskip1.0em frames\_list =scan.grab\_buffer(num\_frames)
    \STATE \hskip1.0em scan.stop\_playing()
    \STATE \hskip1.0em data\_lst= [frame[0].data for frame in frames\_list]
    \STATE \hskip1.0em \textbf{return} pickle.dumps(data\_lst)
  \end{algorithmic}
\end{algorithm}


First, the scan is initialized and enabled to scan a channel \textbf{ch} (lines 2 and 3). Then the scan is triggered (line 6) and waited for a specific time to gather the frames (line 7). This time is calculated by multiplying the number of frames \textbf{num\_frames} with the \textbf{frame time}. Here we added a constant time \textbf{ct} to ensure a proper scanning time is applied to the frames. After that, the frames are stored in \textbf{frames\_list} (line 8), and the scan is stopped (line 9). Finally, the data of the scanned frames are extracted (line 10) and serialized to be wired back to the client (line 11). 

When the client module receives the data, it is deserialized and stored in another data object to be ready for analysis. The serializing and deserializing processes are performed using the pickle python package\cite{pythonpickle}.

\subsection{Probe\_position}
\label{subsec:prob_position}

This task sets the probing position for the scanned data using the coordinates provided by the associated Pyro client application. 
The function for this task is explained below.

\begin{algorithm}[H]
  \begin{algorithmic}[1]
    \Statex \textbf{def} probe\_position(self, x\_coor, y\_coor):
    \STATE \hskip1.0em print(f'probe state:$\backslash$
    \Statex \hskip1.75em {self.stem\_controller.probe\_state}')
    \STATE \hskip1.0em print(f'current position:$\backslash$
    \Statex \hskip1.75em {self.stem\_controller.probe\_position}')
    \STATE \hskip1.0em \textbf{if} x\_coor == y\_coor == 0.0:
    \STATE \hskip1.75em value = None
    \STATE \hskip1.0em \textbf{else}:
    \STATE \hskip1.75em value = Geometry.FloatPoint(y=y\_coor,x=x\_coor)
    \STATE \hskip1.0em  self.stem\_controller.probe\_position = value
  \end{algorithmic}
\end{algorithm}


The task provides the status of the current probing state and the position (lines 1 and 2) before changing the focus to the new position (lines 3-7). 


\bibliographystyle{ieeetr}
\bibliography{./main}

\end{document}